\RequirePackage{fix-cm}
\documentclass[onecollarge]{svjour2}
\smartqed  
\usepackage{graphicx}
\usepackage{amsmath}
\usepackage{slashed}
\usepackage{amstext}
\usepackage{amssymb}
\usepackage{bm}
\usepackage{float}

 \journalname{Few-Body Systems}
\begin{document}

\title{Heavy and heavy-light mesons 
in the Covariant Spectator Theory
}



\author{Alfred Stadler  \and 
	Sofia Leit\~ao\and
          M\,\,T.\,Pe\~na \and
        Elmar P.\,Biernat
}

\authorrunning{ A.\,Stadler, S.\,Leit\~ao, M.\,T.\,Pe\~na, E.\,P.\,Biernat} 

\institute{Alfred Stadler \at
           Departamento de F\'isica, Universidade de \'Evora, 7000-671 \'Evora, Portugal
\\
             \email{stadler@uevora.pt}
	\and
	 Alfred Stadler,  Sofia Leit\~ao, M.\ T.\ Pe\~na and  Elmar P.\ Biernat \at
	CFTP, Instituto Superior T\'ecnico, Universidade de Lisboa, Av. Rovisco Pais 1, 1049-001 Lisboa, Portugal
}

\date{Received: date / Accepted: date}

\maketitle

\begin{abstract}
The masses and vertex functions of heavy and heavy-light mesons, described as quark-antiquark bound states, are calculated with the Covariant Spectator Theory (CST). We use a kernel with an adjustable mixture of Lorentz scalar, pseudoscalar, and vector linear confining interaction, together with a one-gluon-exchange kernel. A series of fits to the heavy and heavy-light meson spectrum were calculated, and we discuss what conclusions can be drawn from it, especially about the Lorentz structure of the kernel. We also apply the Brodsky-Huang-Lepage prescription to express the CST wave functions for heavy quarkonia in terms of light-front variables. They agree remarkably well with light-front wave functions obtained in the Hamiltonian basis light-front quantization (BLFQ) approach, even in excited states.
\end{abstract}

\section{Introduction}
\label{sec:intro}
Early attempts to find a universal description of all known mesons as quark-antiquark states, bound together by a linear and a one-gluon exchange (OGE) potential \cite{Eichten:1978tg,Godfrey:1985}, were very successful. But they were lacking two important aspects: an appropriate treatment of chiral symmetry and of relativity. Modern approaches have to incorporate these features correctly. This paper concentrates on the description of heavy and heavy-light mesons, where the chiral-symmetry aspect is of less importance. Nevertheless, our framework, the Covariant Spectator Theory (CST) \cite{Gross:1969eb,Stadler:2011to}, is able to satisfy the axialvector Ward-Takahashi identity \cite{Gross:1991te,Biernat:2014jt}. Relativistic covariance is one of the cornerstones of this approach, which can be understood as a reorganization of the Bethe-Salpeter equation (BSE) with a complete kernel of ladder and crossed-ladder diagrams. It takes advantage of systematic cancellations between parts of these kernels, which makes it possible to write a three-dimensional equation with a simple kernel that converges faster to the full result than, for instance, the BS ladder approximation.

In contrast to Lattice QCD (e.g., \cite{McNeile:2006,Burch:2006,Dudek:2008,Briceno:2017ve}) and Dyson-Schwinger-Bethe-Salpeter equations (DS-BSE) \cite{Burden:1997vl,Maris:1997,Fischer:2003yf,Eichmann:2009rw,Hilger:2015ty,Eichmann:2016bf}, we implement confinement as originating from an effective linear confining interaction. Another difference is that the former work in Euclidean space, whereas we stay in the physical Minkowski space. This has the advantage that, for instance, meson form factors can be extended quite straightforwardly from space-like to time-like momentum transfer. We are also able to calculate highly excited meson states with about the same effort as ground states, which is not the case for the Euclidean approaches mentioned above.

In this paper, we first review the CST equation for quark-antiquark bound states. Then we present the results of fits to the heavy and heavy-light meson spectrum, with particular emphasis on what can be learned about the Lorentz structure of the confining interaction. Finally, we compare our results with another Minkowski-space approach, the Hamiltonian basis light-front quantization (BLFQ) \cite{Li:2015zda,Li:2017df}.

\section{The one-channel CST bound-state equation}
\label{sec:formalism}

To derive the CST two-body bound equation, we start with the corresponding Bethe-Salpeter equation (BSE) for the vertex function $\Gamma_{BS}(p_1,p_2)$ for an incoming quark with four-momentum $p_2$ (equivalent to an outgoing antiquark with momentum $-p_2$) and an outgoing quark with momentum $p_1$,
\begin{equation}
\Gamma_{BS}(p_1,p_2)= i\int\frac{ {d}^4k}{(2\pi)^4}\,{\cal V}(p,k;P) 
S_1({k}_1)\,\Gamma_{BS}(k_1,k_2)\,S_2(k_2)\,,
\label{eq:BS}
\end{equation}
where ${\cal V}(p,k;P)$ is the two-particle irreducible interaction kernel, which we choose to write as a function of the external and internal relative momenta $p = (p_1+p_2)/2$ and $k = (k_1+k_2)/2$, respectively, and of the total momentum $P=p_1-p_2$.
In general, $S_i(k_i)$ is the dressed propagator of particle $i$ with four-momentum $k_i$. However, in this work we limit ourselves to constituent quarks described with constant masses $m_i$, for which the propagator (with a factor $-i$ removed) simplifies to 
\begin{equation}
S_i(k_i) = \frac{m_i+\slashed{k}_i}{m_i^2-k_i^2-i\epsilon} \, .
\end{equation}
The quark propagators have positive and negative-energy poles at
\begin{equation}
\label{eq:pnpoles}
k_{i0}=\pm(E_{ik} -i\epsilon) \, ,
\end{equation}
where $E_{ik}\equiv \sqrt{m_i^2+{\bf k}_i^2}$. In the rest frame of the bound-state with mass $\mu$, the total momentum is  $P=(\mu,{\bf 0})$. Expressed in terms of the relative-energy component $k_0$, the quark poles of are then located at
\begin{equation}
k_0^{(1\pm)} = \pm (E_{1k} -i\epsilon)-\mu/2 \, , \qquad k_0^{(2\pm)} = \pm (E_{2k} -i\epsilon)+\mu/2 \, .
\end{equation}
In the complex $k_0$ plane,
the positive-energy quark poles, $k_0^{(1+)}$ and $k_0^{(2+)}$, lie in the lower, and the negative-energy quark poles, $k_0^{(1-)}$ and $k_0^{(2-)}$, in the upper half plane.

The idea of CST is to carry out the integration over $k_0$ in (\ref{eq:BS}) by extending the contour into the complex plane and using Cauchy's integral formula, but keeping only the residues of quark propagator poles. This is motivated by partial cancelations between poles in the kernel coming from ladder and crossed-ladder diagrams, which happen in all orders. Leaving out the residues of the kernel's poles can therefore lead to an equation that converges faster to the full BS ladder and crossed-ladder sum than, for instance, the BS ladder approximation \cite{Gross:1969eb}. In contrast to the latter, it also has the correct one-body limit: when one of the particles interacting through the kernel becomes infinitely heavy, the equation turns into an affective one-body (Dirac) equation for the lighter particle in a static field created by the inert source. This is particularly important when dealing with heavy-light systems.

Whenever the residue of a pole in $k_0$ is calculated, the vertex function in the loop is evaluated with one quark momentum on its positive or negative-energy mass shell. This means that we pick up $\Gamma(\hat k_1^+,k_2)$ and $\Gamma(k_1,\hat k_2^+)$ when we close the contour in the lower half plane, and $\Gamma(\hat k_1^-,k_2)$ and $\Gamma(k_1,\hat k_2^-)$ in the upper half plane, where we use the notation $\hat{k}_i^\pm \equiv (\pm E_{ik},{\bf k})$ (note that in the rest frame ${\bf k}_1={\bf k}_2={\bf k}$).

Of course, closing the integration contour in the lower or upper half plane should give the same result, so it doesn't matter where we close the contour. However, when we are leaving out the kernel poles, this is no longer guaranteed. Since a priori no half-plane is preferred, we integrate over $k_0$ by calculating the average over the two half-plane contours. By systematically choosing one of the external momenta as $\hat{p}_i^\pm$, we arrive at a closed set of equations for the CST vertex functions $\Gamma(\hat p_1^\pm,p_2)$ and $\Gamma(p_1,\hat p_2^\pm)$, which we call the ``four-channel covariant spectator equation'' (4CSE), from which we can calculate the complete four-channel CST vertex function
\begin{multline}
 \Gamma_\mathrm{4CSE}(p_1,p_2) =
-\frac 12 \sum_{\eta=\pm}  \Biggl[
 \int \frac{d^3k_1}{(2\pi)^3} \frac{m_1}{E_{1k_1}} 
 \mathcal{V}(p,\hat{k}^\eta_1-P/2)\Lambda_1(\hat{k}^\eta_1)   \Gamma(\hat{k}^\eta_1,\hat{k}^\eta_1-P) S_2(\hat{k}^\eta_1-P) 
 \\
+  \int \frac{d^3k_2}{(2\pi)^3} \frac{m_2}{E_{2k_2}}  
 \mathcal{V}(p,\hat{k}^\eta_2+P/2) 
 S_1(\hat{k}^\eta_2+P) \Gamma(\hat{k}^\eta_2+P,\hat{k}^\eta_2) \Lambda_2(\hat{k}^\eta_2)
\Biggr] \, ,
\label{eq:4CSE}
\end{multline}
where we used $\Lambda_i(\hat{k}) \equiv (m_i+\hat{\slashed{k}})/2m_i$, and $\hat{k}^\eta_i$ means $\hat{k}^+_i$ or $\hat{k}^-_i$.

Another reason for defining $\Gamma_\mathrm{4CSE}$ as in (\ref{eq:4CSE}) is that charge conjugation relates $\Gamma(\hat p_1^+,p_2)$ to $\Gamma(p_1,\hat p_2^-)$, and $\Gamma(p_1,\hat p_2^+)$ to $\Gamma(\hat p_1^-,p_2)$. This implies that one can form vertex functions with definite charge-conjugation parity for equal-flavor quarks.

\begin{figure}[tb]
\begin{center}
\includegraphics[width=0.5\textwidth]{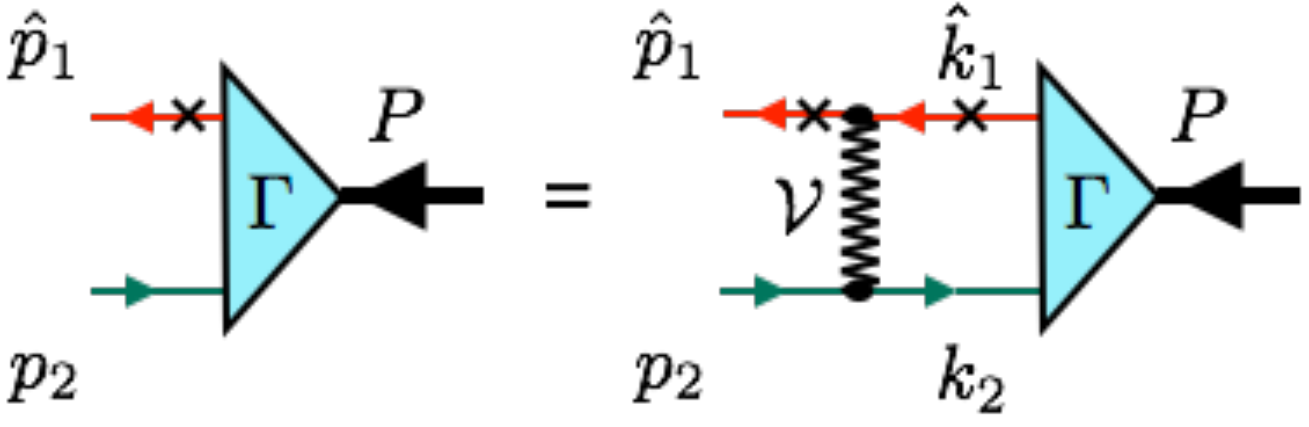}
\caption{The one-channel spectator equation (1CSE) for the bound-state vertex function $\Gamma$ of a quark (particle 1) and an antiquark (particle 2), interacting through a kernel $\cal{V}$. An ``$\times$'' on a line means that the particle is on its positive-energy mass shell, and the corresponding four-momentum carries a ``$\hat{\hspace{10pt}}$''.}
\label{fig:1CSE}
\end{center}
\end{figure}

When the bound-state mass $\mu$ is not small compared to the masses of the constituent quarks, we can use the much simpler one-channel approximation, where we just keep the dominant residue of the positive-energy pole of particle 1 in the lower half-plane (we use the convention that particle 1 is the heavier quark, if the quark masses are different). Since negative-energy poles do no longer occur, we will write simply $\hat p_1$ instead of $\hat p_1^+$. Using a kernel consisting of several terms with different Lorentz structures, 
${\cal V}\equiv \sum_K V_K(p,k)  \Theta_{1}^{K(\mu)} \otimes \Theta^K_{2(\mu)}$, where $\Theta_i^{K(\mu)}={\bf 1}_i, \gamma^5_i,$ or $\gamma_i^\mu$, and where the functions $V_K(p,k)=V_K(\hat{p}_1-P/2,\hat{k}_1-P/2)$ describe the momentum dependence of the kernel labelled $K$,  
the CST vertex function is  given by the so-called one-channel covariant spectator equation (1CSE)
\begin{equation}
\Gamma (\hat{p}_1,p_2)= - \int \frac{d^3k}{(2\pi)^3} \frac{m_1}{E_{1k}} \sum_K V_K(p,k) \Theta_{1}^{K(\mu)}  
\frac{m_1+\hat{\slashed{k}}_1}{2m_1} \Gamma(\hat{k}_1,k_2)
\frac{m_2+\slashed{k}_2}{m_2^2-k_2^2-i\epsilon}\Theta^K_{2(\mu)} \, .
\label{eq:1CSE}
\end{equation}
This approximation is expected to work well for heavy and heavy-light systems, but not for the important case of the pion, where the complete 4CSE should be used. The 1CSE (\ref{eq:1CSE}), shown in diagrammatic form in Fig.~\ref{fig:1CSE}, is manifestly covariant, and has the correct one-body and nonrelativistic limits. However, it is not charge-conjugation symmetric, which means that we cannot assign a $C$-parity to its solutions. This becomes relevant for heavy quarkonia, but, because the splitting between axialvector $C=+$ and $C=-$ pairs is only 5 to 6 MeV in bottomonium and 14 MeV in charmonium, it is of no practical importance for our purposes. Therefore, in this work we sacrifice charge-conjugation symmetry in favor of the much simpler singularity structure of the 1CSE.

The kernel used in our calculations with the 1CSE consists of a covariant generalization of the linear (L) confining potential used in \cite{Leitao:2014}, a one-gluon exchange (OGE), and a covariantized constant (C) interaction,
\begin{equation}
{\cal V}= \left[ (1-y) \left({\bf 1}_1\otimes {\bf 1}_2 + \gamma^5_1 \otimes \gamma^5_2 \right) - y\, \gamma^\mu_1 \otimes \gamma_{2\mu} \right]V_\mathrm{L}(p,k)  +\gamma^\mu_1 \otimes \gamma_{2\mu} \left[ V_\mathrm{OGE}(p,k)+V_\mathrm{C}(p,k) \right]\,.
\label{eq:kernel}
\end{equation}
The Lorentz structure of the confining interaction is not precisely known. We choose a scalar-plus-pseudoscalar mixed with vector structure, controlled by a mixing parameter $y$ that we will determine by fitting to the data. In principle, scalar and pseudoscalar interactions break chiral symmetry. However, we have shown that the axial-vector Ward-Takahashi identity can by satisfied when our linear confining interaction has  equal-weight scalar and pseudoscalar components \cite{Biernat:2014uq}. 

The momentum dependence of the interaction kernel in the 1CSE is chosen to depend only on the momentum transfer, $q=p-k=\hat{p}_1-\hat{k}_1$,
\begin{equation}
V_\mathrm{L}(p,k)  = -8\sigma \pi\left[\left(\frac{1}{q^4}-\frac{1}{\Lambda^4+q^4}\right)-\frac{E_{1p}}{m_1}(2\pi)^3 \delta^3 (\mathbf{q})\int \frac{d^3 k'}{(2\pi)^3}\frac{m_1}{E_{1k'}}\left(\frac{1}{q'^4}-\frac{1}{\Lambda^4+q'^4}\right)\right],
\end{equation}
\begin{equation}
V_\mathrm{OGE}(p,k)  = -4 \pi \alpha_s \left(\frac{1}{q^2}-\frac{1}{q^2-\Lambda^2}\right), \quad
V_\mathrm{C}(p,k) = (2\pi)^3\frac{E_{1k}}{m_1} C \delta^3 (\mathbf{q})\,,
\label{eq:V}
\end{equation}
where $q^{\prime}=p-k^{\prime}=\hat{p}_1-\hat{k}'_1$.

The three coupling strengths, $\sigma$, $\alpha_s$, and $C$, are free parameters of the model, and we use Pauli-Villars regularization for both the linear and the OGE kernels, which yields one additional cut-off parameter $\Lambda$. Here we don't treat $\Lambda$ as adjustable, but we scale it with the heavy mass, in the form $\Lambda=2 m_1$. The signs in (\ref{eq:kernel}) are chosen such that the nonrelativistic limit of the kernel yields, for any value of $y$, always the same Fourier transform of the Cornell potential $V(r)=\sigma r -\alpha_s/r-C$.

In order to solve Eq.\ (\ref{eq:1CSE}), we make use of the---frame-dependent---decomposition
\begin{equation}
\frac{m_2+\slashed{k}_2}{m_2^2-k_2^2-i\epsilon} 
=
\frac{m_2}{E_{2k}} \sum_{\rho_2,\lambda_2} 
\rho_2 \frac{u_2^{\rho_2}({\bf k},\lambda_2) \bar{u}_2^{\rho_2}({\bf k},\lambda_2)}{E_{2k}-\rho_2 k_{20}-i\epsilon} \, ,
\label{eq:propdecomp}
\end{equation}
as well as of
\begin{equation}
\frac{m_1+\hat{\slashed{k}}_1}{2m_1} = \sum_{\lambda_1} u_1^+({\bf k},\lambda_1) \bar{u}_1^+({\bf k},\lambda_1) \, ,
\label{eq:onshellproj}
\end{equation}
where the $\rho$-spinors $u_i^\pm$ with helicity $\lambda$ are defined as
\begin{equation}
u_i^+({\bf k},\lambda) \equiv u_i({\bf k},\lambda) \, , \qquad u_i^-({\bf k},\lambda) \equiv v_i(-{\bf k},\lambda) \, .
\end{equation}
Next we define CST wave functions when quark 1 is on the positive mass shell,
\begin{align}
 \Psi_{\lambda_1 \lambda_2}^{+\rho_2} ({\bf p}) 
 & \equiv 
 \sqrt{\frac{m_1m_2}{E_{1p}E_{2p}}}
  \frac{ \bar u_1^{+}({\bf p}, \lambda_1) \Gamma (\hat p_1,p_2) u_2^{\rho_2}(\rho_1{\bf p},\lambda_2)}
 {\rho_2 E_{2p}  - E_{1p}+\mu -i\epsilon}  \, ,
 \label{eq:CSTwfs}
\end{align}
and the spinor matrix elements of the interaction vertices,
\begin{align}
\Theta^{K,\rho\rho'}_{i,\lambda\lambda'}({\bf p},{\bf k}) & \equiv 
\bar{u}_i^{\rho}({\bf p},\lambda)\Theta^K_i  u_i^{\rho'}({\bf k},\lambda')
\, .
\label{eq:thetaO}
\end{align}
Equation (\ref{eq:1CSE}) can now be rewritten 
as the 1CSE for the CST wave function
\begin{multline}
  ( \rho_2  E_{2p}-  E_{1p}+\mu) \Psi_{\lambda_1 \lambda_2}^{+\rho_2} ({\bf p}) 
 =
- \hspace{-3mm} \sum_{K  \lambda'_1 \lambda'_2  \rho'_2 } \int \frac{\mathrm d^3 k}{(2\pi)^3}N_{12}(p,k) 
V_K({\bf p},{\bf k}) 
 \Theta^{K,++}_{1,\lambda_1\lambda'_1}({\bf p},{\bf k}) 
  \Psi_{\lambda'_1 \lambda'_2}^{+\rho'_2} ({\bf k})\\
 \hspace{6mm} \times 
  \Theta^{K,\rho'_2\rho_2}_{2,\lambda'_2\lambda_2}({\bf k},{\bf p}) \, , 
  \label{eq:1CSEwf}
\end{multline}
where we have introduced the shorthand
$N_{12}(p,k) \equiv m_1 m_2 /\sqrt{E_{1p} E_{2p}E_{1k} E_{2k}}$. Note that, in the kinematics of the 1CSE, the functions $V_K$ depend only on the three-vectors $\bf p$ and $\bf k$.

These wave functions are normalized according to
\begin{equation}
2\mu=N_c \sum_{\lambda_1\lambda_2 \rho_2}
\int \frac{d^3 k}{(2\pi)^3}  \left[\Psi_{\lambda_1 \lambda_2}^{+\rho_2} ({\bf k})\right]^\dagger  \Psi_{\lambda_1 \lambda_2}^{+\rho_2} ({\bf k}) \, ,
\end{equation}
where $N_c=3$ is the number of colors.
 We expand the CST wave functions in a basis of angular wave functions $K_j^{\rho_2}(\hat{\bf p})$ with definite orbital angular momentum $L$ and total quark-antiquark spin $S$, and the corresponding radial wave functions $\psi_j^{\rho_2}(p)$,
\begin{equation}
\Psi^{+\rho_2}_{\lambda_1\lambda_2} ({\bf p})=\sum_j \psi_j^{\rho_2}(p) \chi^\dagger_{\lambda_1}(\hat{\bf p})\, K_j^{\rho_2}(\hat{\bf p}) \, \chi_{\lambda_2}(\hat{\bf p}).
\label{eq:PSI}
\end{equation}
\begin{table*}[tb]
\begin{center}
    \caption{Summary table of the kernel parameters of the different fitting models considered in this work (we use $m_u=m_d \equiv m_q$). $N_\mathrm{st}$ is the number of states in the data set used in fitting the model. $\delta_\mathrm{rms}$ indicates the minimized root mean square difference with respect to the data set used in the fit, and $\Delta_\mathrm{rms}$ is the root mean square difference with respect to data set S3, including both fitted and predicted states. The values in boldface were held fixed. The units for the quark masses, $\delta_\mathrm{rms}$, $\Delta_\mathrm{rms}$, and $C$ are GeV, and $\sigma$ is in GeV$^2$.     \label{tab:parameters}}
\begin{tabular}{  c   c c c c cccc c c c  }
\hline\noalign{\smallskip}
  Model & $\sigma$  & $\alpha_s$ &$C$  & $y$ & $m_b$ & $m_c$  &$m_s$ &$m_q$ & $N_\mathrm{st}$ & $\delta_\mathrm{rms}$  & $\Delta_\mathrm{rms}$ \\ 
  \noalign{\smallskip}\hline\noalign{\smallskip}
  M0$_{\text{S1}}$   &     0.2493 & 0.3643 & 0.3491 & {\bf 0.0000} & {\bf 4.892}& {\bf 1.600} &{\bf 0.4478} & {\bf 0.3455} & 9 & 0.017& 0.037\\
    M1$_{\text{S1}}$& 0.2235 & 0.3941 & 0.0591  &0.0000 & 4.768 & 1.398& 0.2547& 0.1230& 9  & 0.006 &0.041 \\
    M0$_{\text{S2}}$&  0.2247 & 0.3614 & 0.3377  & {\bf 0.0000} & {\bf 4.892} & {\bf 1.600} & {\bf 0.4478}& {\bf 0.3455} & 25  &0.028 &0.036 \\
M1$_{\text{S2}}$&  0.1893 & 0.4126 & 0.1085  &0.2537 & 4.825 & 1.470& 0.2349& 0.1000 & 25  &0.022 & 0.033\\
M1$_{\text{S2}'}$& 0.2017 & 0.4013 & 0.1311 & 0.2677 &  4.822 & 1.464 & 0.2365 & 0.1000 &  24& 0.018 & 0.033 \\
M1$_{\text{S3}}$& 0.2022 & 0.4129 & 0.2145  &0.2002 & 4.875 & 1.553 & 0.3679 & 0.2493 & 39  &0.030 & 0.030 \\
M0$_{\text{S3}}$&  0.2058 & 0.4172 & 0.2821  &{\bf 0.0000}  & 4.917 & 1.624 & 0.4616 & 0.3514 & 39& 0.031 & 0.031\\
\noalign{\smallskip}\hline
    \end{tabular}
\end{center}
    \end{table*}
%
\begin{figure}[h!]
\begin{center}
  \includegraphics[width=1.\textwidth]{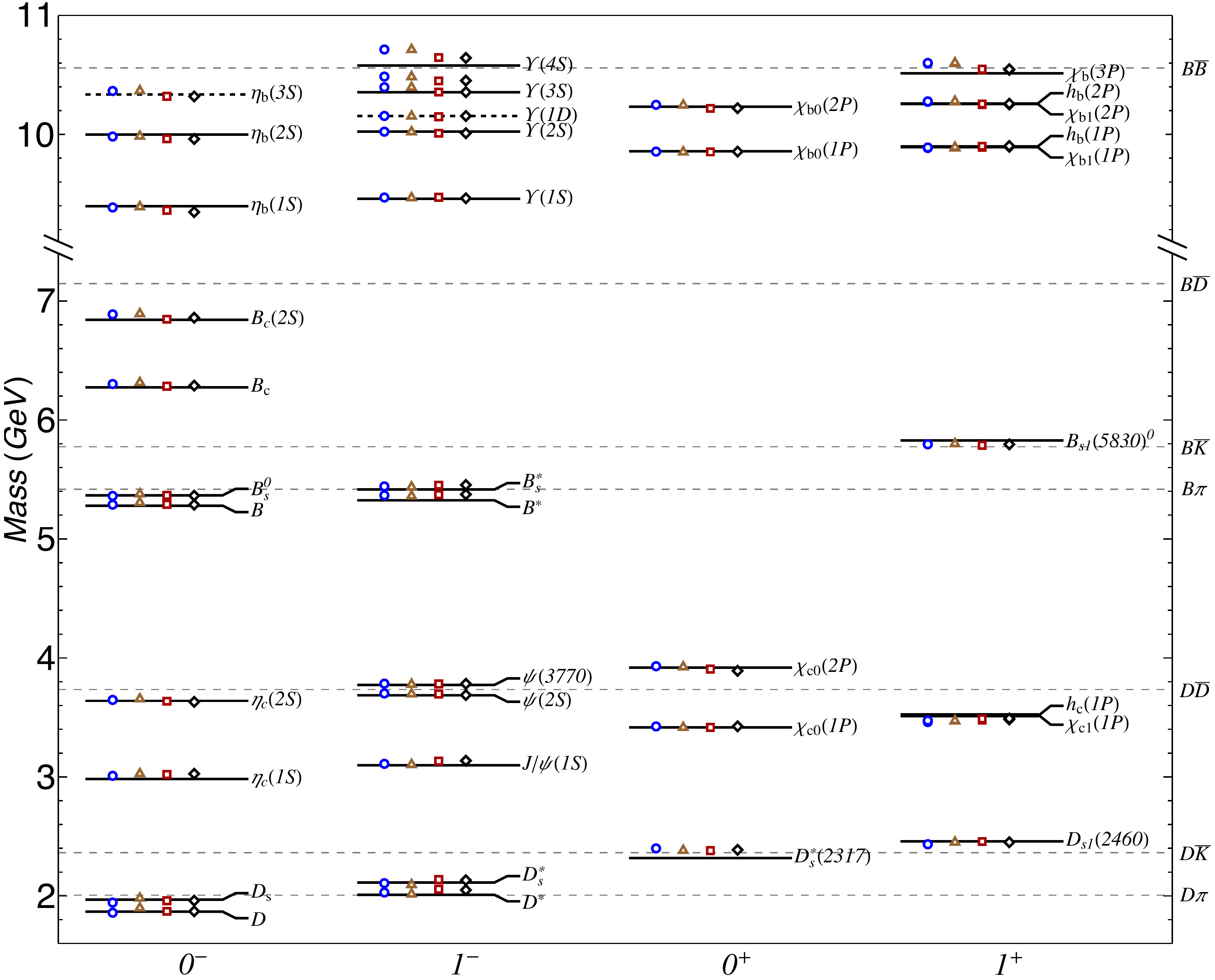}
\caption{Spectrum of heavy and heavy-light mesons with $J^P=0^\pm$ and $1^\pm$. The symbols represent calculations with models M0$_{\text{S1}}$ (circles), M0$_{\text{S1}}$ with PS coupling turned off (triangles), M1$_{\text{S3}}$ (squares), and M0$_{\text{S3}}$ (diamonds). The solid lines are experimental data from the PDG \cite{PDG2014}.}
\label{fig:spectrum}       
\end{center}
\end{figure}
The angular functions can be found in \cite{Leitao:2017yu}, and $\chi_\lambda(\hat{\bf p})$ is a two-component spinor with helicity $\lambda$, for a three-momentum pointing into the direction $\hat{\bf p}$.

These wave functions contain relativistic components not present in nonrelativistic solutions. For instance, pseudoscalar states are dominated by $S$-waves, but they are coupled to small $P$-waves (the opposite parity of their orbital wave function is compensated by an opposite intrinsic parity of one of the quarks, leaving the total parity unchanged) that vanish in the nonrelativistic limit. In the case of vector mesons, coupled $S$- and $D$-waves are accompanied by relativistic spin-singlet  and spin-triplet  $P$-waves, denoted $P_s$ and $P_t$, respectively.

\section{Numerical results and conclusions}

Substitution of (\ref{eq:PSI}) in Eq.\ (\ref{eq:1CSEwf}) leads to a system of coupled equations for the radial wave function components, which in the case of the 1CSE has the form of a linear eigenvalue problem. We solve it numerically by representing the radial wave functions in a basis of cubic B-splines, adjusted for a correct asymptotic behavior.

Our global model parameters (Tab.~\ref{tab:parameters}) were determined through least square fits to various sets of experimental masses for $J^P=0^\pm$ and $1^\pm$ mesons. The set S1 consists of 9 pseudoscalar (PS) states, S2 adds scalar (S) and vector (V) mesons to a total of 25 states (the set S2$^\prime$ leaves out the highly excited $\Upsilon(4S)$), and S3 includes axialvector (AV) mesons with a total of 39 states (a detailed list of these data sets can be found in \cite{Leitao:2017yu}). The mass spectra for some cases are shown in Fig.~\ref{fig:spectrum}, the model parameters and corresponding rms differences between calculated masses and experimental data are shown in Tab.~\ref{tab:parameters}. Also shown in Fig.~\ref{fig:spectrum} is that---as expected---the results remain almost unchanged when the PS coupling in the confining kernel of model M0$_{\text{S1}}$ is turned off (the largest difference is about 40 MeV in PS $c\bar{q}$).

\begin{table}[tb]
\begin{center}
\caption{Masses (in GeV) of the lowest four states (numbered by $n$) of selected pseudoscalar and vector mesons (``$q$'' is a $u$ or $d$ quark), calculated with model M1$_{\text{S3}}$, and using different numbers of splines in the expansion of the radial wave functions. }
\label{tab:convergence}       
\begin{tabular}{cccrrrrr}
\hline\noalign{\smallskip}
 & & & \multicolumn{5}{c}{Number of splines}\\
Meson & $J^P$ &$n$ & \multicolumn{1}{c}{12} & \multicolumn{1}{c}{24} & \multicolumn{1}{c}{36} & \multicolumn{1}{c}{48} & \multicolumn{1}{c}{64}  \\
\noalign{\smallskip}\hline\noalign{\smallskip}
$b\bar{b}$ & $0^-$& 1 & 9.37765 & 9.37886 & 9.37917 & 9.37931 & 9.37940 \\
 &  & 2 &9.96915 & 9.96932 & 9.96938 & 9.96939 & 9.96939 \\
 &  & 3 &10.33061 & 10.32623 & 10.32623 & 10.32622 & 10.32621 \\
 &  & 4 &10.61822 & 10.61660 & 10.61646 & 10.61643 & 10.61641 \\
\noalign{\smallskip}
$b\bar{b}$ & $1^-$& 1 & 9.47414  &  9.47411 & 9.47409 & 9.47407 & 9.47406  \\
 &  & 2 & 10.01186 & 10.01147 & 10.01141 & 10.01138 & 10.01135  \\
 &  & 3 & 10.14699 & 10.14692 & 10.14702 &  10.14714 & 10.14731 \\
 &  & 4 & 10.36325 & 10.35767 & 10.35758 & 10.35755 & 10.35751 \\
\noalign{\smallskip}
$c\bar{c}$ & $0^-$& 1 & 3.02240 & 3.02341 & 3.02380 & 3.02400 & 3.02414 \\
 &  & 2 & 3.63778 & 3.63814& 3.63832 & 3.63843 & 3.63850  \\
 &  & 3 & 4.09893 & 4.09910 & 4.09925 &  4.09933 & 4.09938 \\
 &  & 4 & 4.49972 & 4.49926 & 4.49940 & 4.49947 & 4.49952 \\
\noalign{\smallskip}
$c\bar{c}$ & $1^-$& 1 &  3.13139 & 3.13154 & 3.13163 & 3.13169 & 3.13174  \\
 &  & 2 & 3.69834 & 3.69840& 3.69847 & 3.69853 & 3.69857 \\
 &  & 3 & 3.75095 & 3.75366 & 3.75659 & 3.75966 &  3.76395 \\
 &  & 4 & 4.14245 & 4.14248 & 4.14257 & 4.14263 & 4.14267 \\
\noalign{\smallskip}
$c\bar{q}$ & $0^-$& 1 & 1.86997 & 1.87122 & 1.87182 & 1.87217 &  1.87247 \\
 &  & 2 & 2.51166 & 2.51196& 2.51213 & 2.51227 & 2.51242  \\
 &  & 3 & 2.99045 & 2.99065 & 2.99071 & 2.99079 & 2.99090 \\
 &  & 4 & 3.40197 & 3.40221 & 3.40225 & 3.40232 & 3.40241 \\
\noalign{\smallskip}
$c\bar{q}$ & $1^-$& 1 &  2.05555 & 2.05597 & 2.05612 & 2.05620 & 2.05626 \\
 &  & 2 & 2.61323 & 2.61365 & 2.61383 & 2.61397 & 2.61411 \\
 &  & 3 & 2.65564 & 2.65763 & 2.66005 & 2.66273 &  2.66654 \\
 &  & 4 & 3.06017 & 3.06073 & 3.06096 & 3.06115 & 3.06135 \\
\noalign{\smallskip}\hline
\end{tabular}
\end{center}
\end{table}

\begin{figure}[hbt]
  \includegraphics[width=1.0\textwidth]{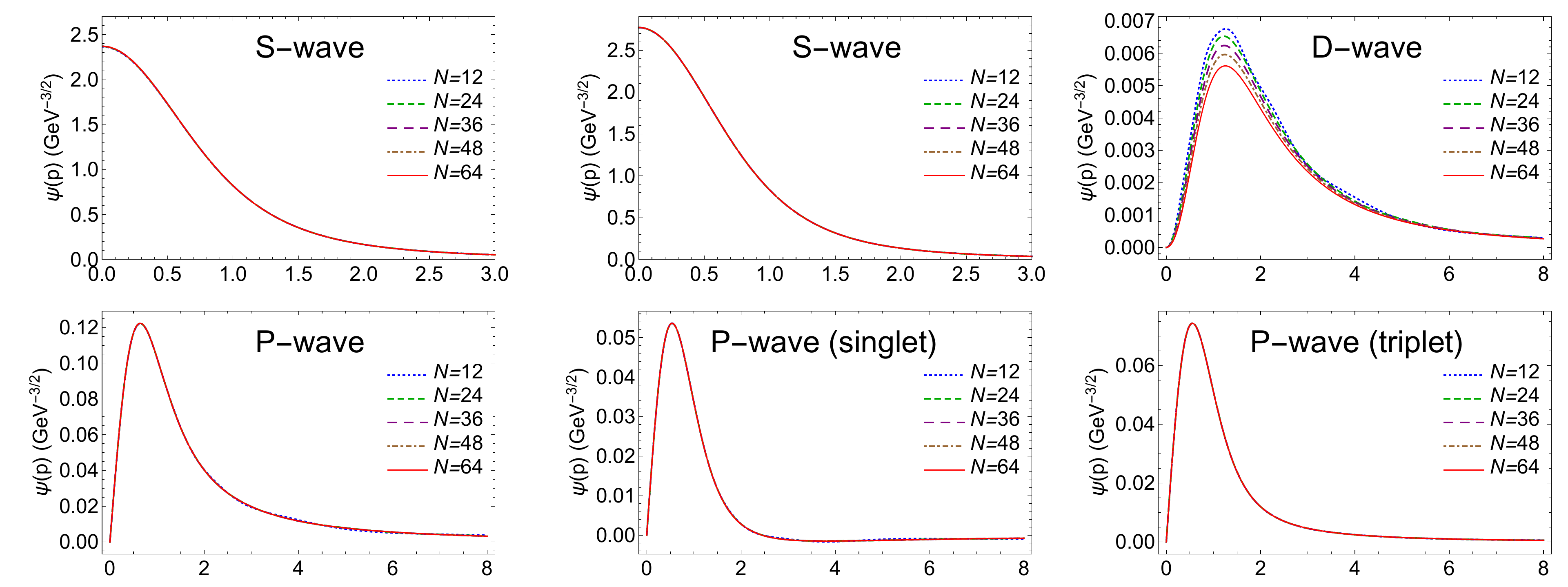}
\caption{Wave function components for the ground states of PS (left column) and V charmonium with model M1$_{\text{S3}}$, calculated with different numbers of splines. In each case, the five curves coincide almost everywhere.}
\label{fig:wfs_SN}       
\end{figure}

The models M0$_{\text{S1}}$ and  M0$_{\text{S2}}$ (identical to P1 and PSV1 of Ref.~\cite{Leitao:2017it}) where fitted for $y=0$ (no Lorentz vector coupling in the confining kernel) and keeping the quark masses fixed. It is remarkable that a fit to a few PS states alone is already sufficient to predict the spectrum of V, S, and AV mesons with very good quality. In a subsequent work \cite{Leitao:2017yu}, we allowed $y$ and the quark masses to vary as well, the latter being a rather challenging task with respect to the required computing time. It turned out that, depending on the data set used in the fit, $y$ takes on values different from zero. The fit M1$_{\text{S1}}$ still prefers $y=0$, whereas M1$_{\text{S3}}$, the model with the best overall fit, yields $y=0.20$. However, more detailed studies showed that the minimum of the least-square-difference at $y=0.20$ is very shallow, and fixing $y$ anywhere between $0$ and $0.3$ gives fits of essentially the same quality. One can see this, for instance, by comparing M1$_{\text{S3}}$ and M0$_{\text{S3}}$ in Fig.~\ref{fig:spectrum} and Tab.~\ref{tab:parameters}: M0$_{\text{S3}}$ is obtained with a fixed $y=0$, and its rms difference to the data is only marginally worse than M1$_{\text{S3}}$'s. We can conclude that the mass spectrum alone does not provide very tight constraints on the Lorentz mixing parameter $y$, and we have to look for other observables in order to obtain more detailed information on the Lorentz structure of the confining interaction.
Similarly, we found that a relatively broad range of constituent quark masses is compatible with a good description of the mass spectrum.

The models of Tab.~\ref{tab:parameters} were all fitted in a basis of 12 splines. Meanwhile we were able to improve our computational methods, which allows us to perform these calculations in much larger basis spaces. Table~\ref{tab:convergence} demonstrates that we obtain excellent numerical stability, and in most cases the result with 12 splines is already converged at the 1 MeV level. Figure \ref{fig:wfs_SN} shows that also the wave functions are very stable.

\begin{figure}[b]
\begin{center}
  \includegraphics[width=0.4\textwidth]{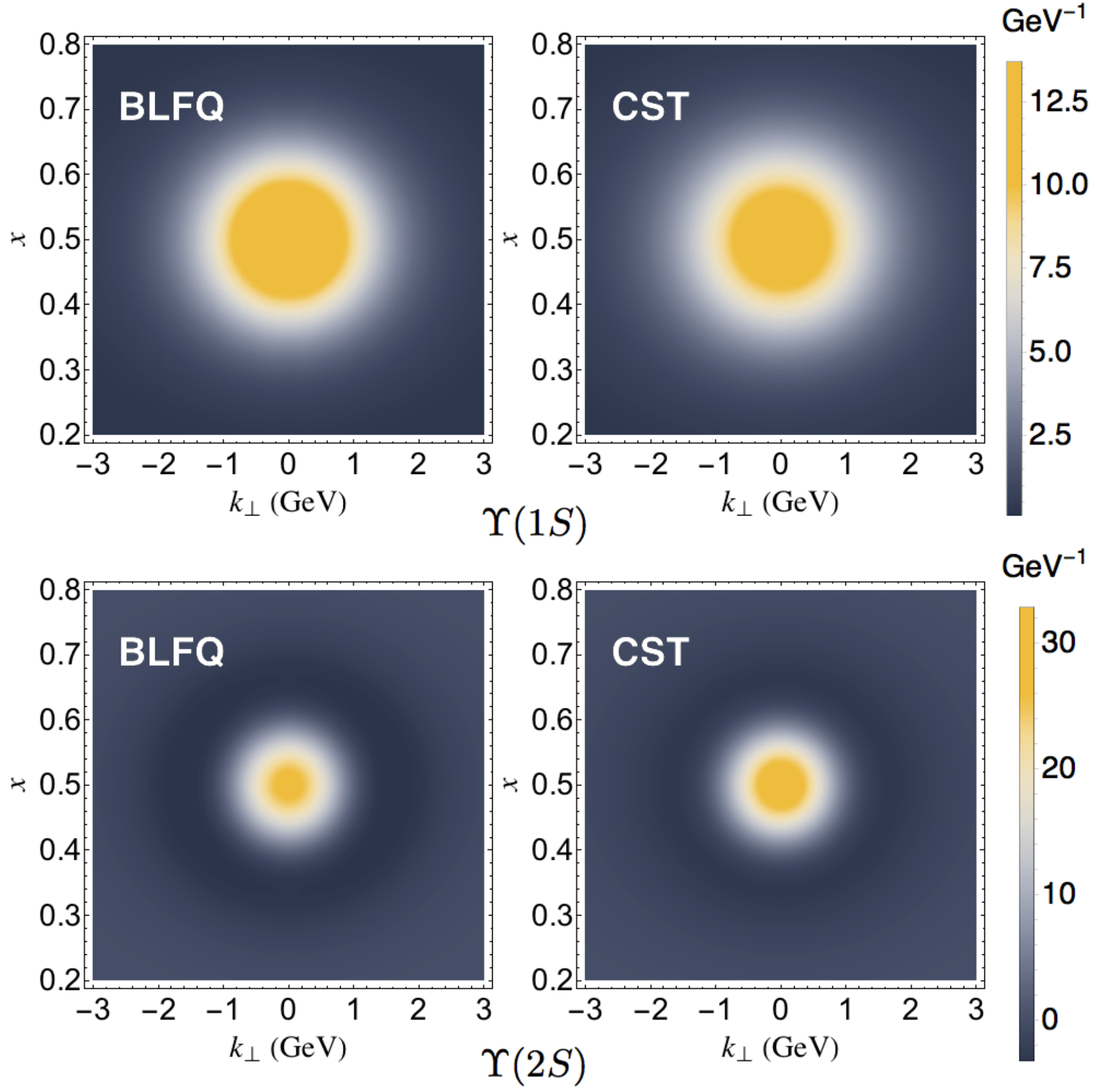}\hspace{2mm}
    \includegraphics[width=0.4\textwidth]{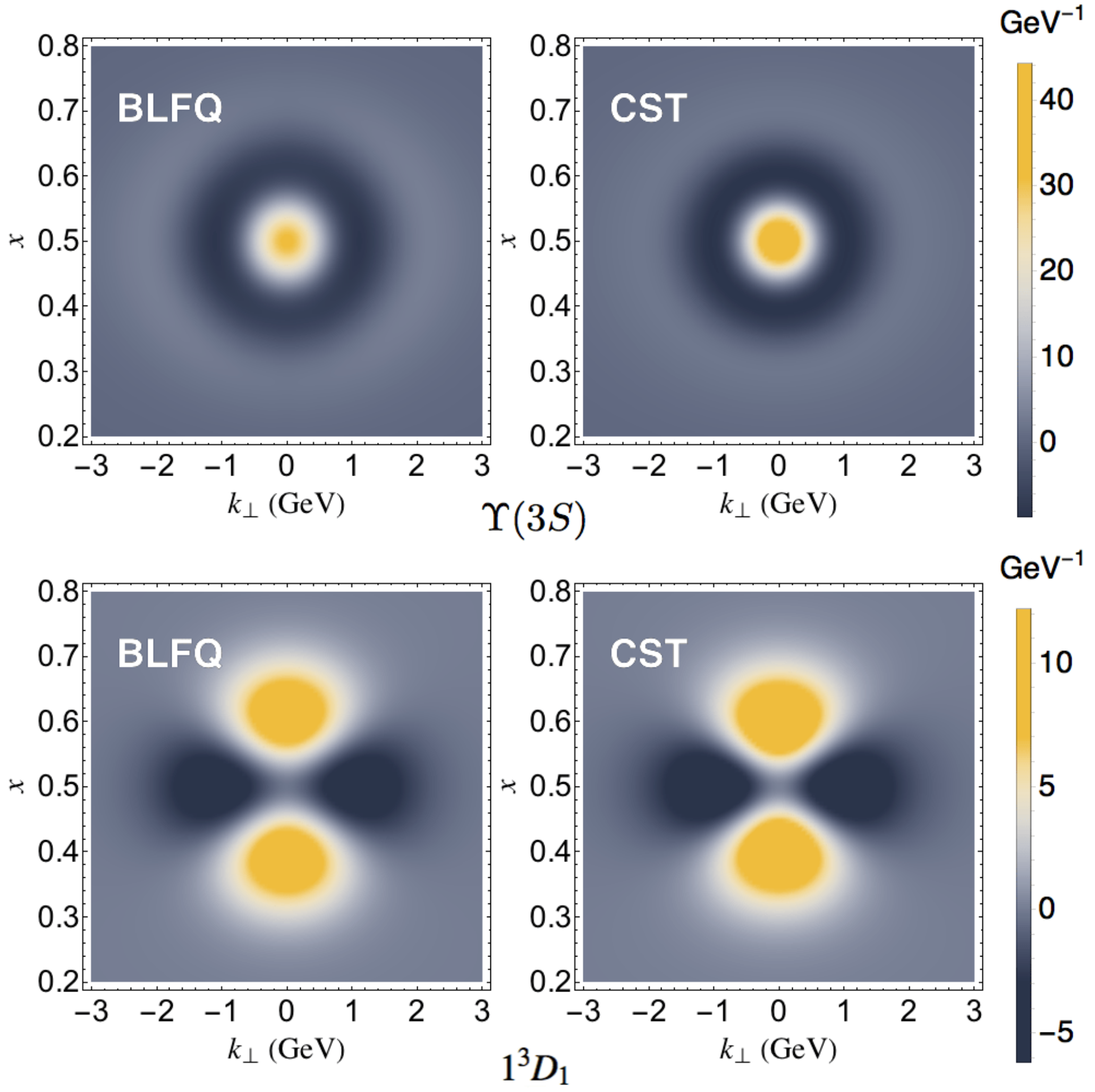}
\caption{Dominant triplet components  of the BLFQ- and CST-LFWFs for the four lowest vector bottomonium states.}
\label{fig:LFWFs}       
\end{center}
\end{figure}

Another Minkowski-space approach to heavy quarkonia is the Hamiltonian basis light-front quantization (BLFQ), which describes the quarkonium mass spectra with similar quality as do our CST models \cite{Li:2015zda,Li:2017df}. It would be interesting to compare the two approaches also at the level of wave functions. However, it is not yet clear how to project our CST wave functions onto the light front in a rigorous manner. The main difficulty is that re-writing the CST wave functions in terms of light-cone variables and boosting to the infinite-momentum frame does not eliminate all components with a longitudinal momentum fraction $x$ of the quark larger than 1. For now, we circumvented this problem by applying the often-used Brodsky-Huang-Lepage (BHL) prescription \cite{Bro1983} to produce approximate CST light-front wave functions (LFWFs). As an example, Fig.~\ref{fig:LFWFs} shows the dominant triplet component of the BLFQ- and CST-LFWFs for the four lowest vector bottomonium states. The two sets of wave functions look very similar, even in higher excited states. Differences become visible only in the subdominant wave function components \cite{Leitao:2017xi}. When these LFWFs are used to calculate leading-twist parton distribution amplitudes and parton distribution functions, we again find that they are consistent with each other \cite{Leitao:2017xi}. Considering that we are comparing two very different frameworks and with different dynamics (BLFQ is rooted in light-front holographic QCD), it is surprising that the requirement to reproduce the mass spectrum comparably well is sufficient to lead to such a remarkable agreement of the wave functions. 

This first attempt to find connections between two different approaches has already led to interesting results. In future work, we will try to replace the BHL prescription by a more rigorous method of projecting the CST wave functions onto the light front. Also planned is an extension of the comparison with BLFQ to lighter systems.

\begin{acknowledgements}
This work was supported by Funda\c c\~ao para a Ci\^encia e a 
Tecnologia (FCT) under Grants No. CFTP-FCT (UID/FIS/00777/2013),  No. SFRH/BPD/100578/2014, and No. SFRH/BD/92637/2013.
\end{acknowledgements}



\end{document}